\documentclass[10pt,aps,prd,twocolumn,showpacs,superscriptaddress,eqsecnum,longbibliography,nofootinbib]{revtex4-1}
\usepackage{mathrsfs,amsmath,amsthm,latexsym,amssymb,amsfonts,epsfig,cancel}
\usepackage{xcolor}
\definecolor{navy}{RGB}{0,0,150}
\usepackage[colorlinks, linkcolor=navy, citecolor=navy, urlcolor=navy, plainpages=false, pdfstartview=FitH]{hyperref}
\usepackage{appendix}
\newcommand{\figref}[1]{Fig.~\ref{#1}}
\allowdisplaybreaks
\newcommand{\GZU}{School of Physics, Guizhou University, Guiyang 550025, China}
\newcommand{\UOW}{Faculty of Physics, University of Warsaw, Pasteura 5, 02-093 Warsaw, Poland}
\newcommand{\BNU}{Department of Physics, Beijing Normal University, Beijing 100875, China}

\begin{document}

\title{Loop quantum cosmology from an alternative Hamiltonian. II. \\Including the Lorentzian term}

\author{Jinsong Yang}
\email{jsyang@gzu.edu.cn}
\affiliation{\GZU}

\author{Cong Zhang}
\email{Cong.Zhang@fuw.edu.pl}
\affiliation{\UOW}

\author{Yongge Ma}
\thanks{Corresponding author}
\email{mayg@bnu.edu.cn}
\affiliation{\BNU}


\begin{abstract} 

The scheme of using the Chern-Simons action to regularize the gravitational Hamiltonian constraint is extended to including the Lorentzian term in the $k=0$ cosmological model. The Euclidean term and the Lorenzian term are thus regularized separately mimicking the treatment of full loop quantum gravity. The new quantum dynamics for the spatially flat Friedmann-Robertson-Walker model with a massless scalar field as an emergent time is studied. By semiclassical analysis, the effective Hamiltonian constraint is obtained, which indicates that the new quantum dynamics has the correct classical limit. The classical big-bang singularity is again replaced by a quantum bounce. Similar to the case of quantizing only the Euclidean term, the backward evolution of the cosmological model determined by the new effective Hamiltonian will be bounced to an asymptotic de Sitter universe coupled to a massless scalar field, while the problem of negative energy density of matter in the former case is resolved.

\end{abstract}


\maketitle

\section{Introduction}

The spacetime singularities, including the cosmological big-bang/big-crunch singularity and the interior black hole singularity, imply that general relativity (GR) fails in giving reliable physical predicts when spacetime curvature increases unboundedly. Hence some quantum theory of gravity should take over the classical theory to solve the singularity problem and make the physical predicts there. How to incorporate the principles of GR and quantum theory into the ultimate theory of quantum gravity in a consistent way remains a huge challenge in modern theoretical physics. The hunt for quantum gravity has bred a number of theories.  Among these theories, loop quantum gravity (LQG), as a nonperturbative and background-independent approach to the quantization of GR has made remarkable achievements in the past thirty years (see \cite{Rovelli:2004tv,Thiemann:2007pyv} for books, and \cite{Thiemann:2002nj,Ashtekar:2004eh,Han:2005km,Giesel:2012ws,Ashtekar:2017awx} for articles). LQG naturally predicts that the classical continue spacetime breaks down at Planck scale, while the discrete spacetime geometry appears. All the geometric operators corresponding to the classical length, area and volume have the discrete spectra \cite{Rovelli:1994ge,Ashtekar:1996eg,Ashtekar:1997fb,Yang:2016kia,Thiemann:1996at,Ma:2010fy}. The ADM energy of an asymptotically flat spacetime and various expressions of quasi-local energy have been well defined as operators in LQG based on different physical considerations \cite{Thiemann:1997rs,Major:1999xu,Yang:2008th}. Algebraic and graphical methods of calculus were also developed for LQG in order to investigate the properties of physically interested operators, and to implement some consistency checks on different regularization schemes for quantum operators \cite{DePietri:1996tvo,Thiemann:1996au,Brunnemann:2004xi,Brunnemann:2005ip,Borissov:1997ji,Yang:2015wka,Giesel:2005bk,Giesel:2005bm,Yang:2019xms}. The ideas and techniques developed in LQG has also been applied to high-dimensional GR \cite{Bodendorfer:2011nx,Long:2019nkf,Long:2020wuj}, as well as the $f(R)$, the scalar-tensor, and the Weyl theories of gravity \cite{Zhang:2011vi,Zhang:2011qq,Zhang:2011vg,Zhang:2011gn,Ma:2011aa,Zhou:2012ie,Han:2013noa,Chen:2018dqz}. Despite of these achievements, how to implement the quantum dynamics of LQG is still an open issue. Recently, some new strategies inspired by that in \cite{Thiemann:1996aw,Thiemann:1997rt} to implement the quantization for the Hamiltonian constraint have been carried out in \cite{Yang:2015zda,Alesci:2015wla}, and then the resulting operators were studied in detail in \cite{Alesci:2011ia,Thiemann:2013lka,Zhang:2018wbc,Zhang:2019dgi}. Moreover, some constructions of the Hamiltonian constraint operator displaying a nontrival anomaly-free representation of the classical constraint algebra in a certain sense were studied in \cite{Tomlin:2012qz,Varadarajan:2012re,Varadarajan:2018tei,Varadarajan:2019wpu}. Beside these progresses, the Chern-Simons action was employed to construct the Hamiltonian constraint of GR in \cite{Soo:2005gw,Soo:2007hj}. This provides a new regularization strategy in LQG, which deserves further studying. 

To test the ideas and techniques employed in LQG, the symmetry-reduced models of LQG, including loop quantum cosmology (LQC) and loop quantum black holes, have been studied \cite{Bojowald:2001xe,Ashtekar:2003hd,Ashtekar:2005qt}. In particular, the classical big-bang singularity in the homogenous and isotropic model of cosmology is resolved in certain sense in the framework of LQC \cite{Bojowald:2001xe,Ashtekar:2003hd,Ashtekar:2006rx,Ashtekar:2006uz,Ashtekar:2006wn,Ding:2008tq,Yang:2009fp,Assanioussi:2018hee,Li:2018opr}. The interior of Schwarzschild black holes is also quantized by applying LQC techniques, leading to the singularity resolution \cite{Ashtekar:2005qt,Ashtekar:2018lag,Ashtekar:2018cay,Zhang:2019acn} and black hole remnant \cite{Zhang:2020qxw}. Moreover, the LQC framework is also extended to the cosmological models of scalar-tensor theories of gravity \cite{Zhang:2012em,Zhang:2012ta,Artymowski:2013qua,Han:2019mvj,Song:2020pqm}. All these achievements are related to the implementation of quantum dynamics in the symmetry-reduced models, inheriting from the full LQG. In contrary to LQG where the spatial diffeomorphisms plays a crucial role in removing the regulator of the regularized Hamiltonian constraint in the quantization procedure \cite{Rovelli:2004tv,Thiemann:1996aw,Thiemann:2007pyv,Thiemann:2002nj,Ashtekar:2004eh,Han:2005km}, the gauge degrees of freedom of spatial diffeomorphisms have been fixed in LQC. Hence, removing the regulator from the regularized Hamiltonian constraint operator requires more careful attention. To quantize the gravitational Hamiltonian constraint in the strategy of LQG, one needs to express the curvature of the gravitational connection in terms of holonomies around loops. In LQC, one shrinks the area of a loop to the smallest nonzero area eigenvalue, known as the area gap in LQG. There are different ways to implement this strategy, including the so-called $\mu_0$-scheme \cite{Ashtekar:2006uz} and the $\bar{\mu}$-scheme \cite{Ashtekar:2006wn}. In the physically more reasonable $\bar{\mu}$-scheme in which the quantum bounce occurs only when the matter density reaches Planck scale, one shrinks the loop so that its physical area, rather than its fiducial area, reaches the area gap, and hence the regulator becomes a dynamical variable.

In LQG, the gravitational Hamiltonian constraint consists of the Euclidean term and the Lorentzian term \cite{Rovelli:2004tv,Thiemann:2007pyv,Thiemann:2002nj,Ashtekar:2004eh,Han:2005km,Giesel:2012ws}. In the spatially flat and homogeneous models of the classical theory, the two terms are proportional to each other. Hence one used to quantize the Euclidean term, leading to a symmetric bounce for the LQC model with a massless scalar field \cite{Ashtekar:2006rx,Ashtekar:2006uz,Ashtekar:2006wn,Ding:2008tq}. However, to inherit more features from full LQG, one can also treat the two terms independently \cite{Yang:2009fp}, leading to an asymmetric bounce \cite{Assanioussi:2018hee,Li:2018opr}. Recently, it was shown that the effective Hamiltonian of one of the models in the latter treatment proposed in \cite{Yang:2009fp} could be reproduced by a suitable semiclassical analysis of Thiemann's Hamiltonian in full LQG \cite{Dapor:2017rwv,Han:2019feb}. Notice that all of the above results were based on the regularization scheme employing the Thiemann's trick. Had one adopted a different scheme, the quantization of the Euclidean term could also lead to an asymmetric bounce \cite{Liegener:2019zgw,Yang:2019ujs}.

In particular, the Chern-Simons action was employed to construct an alternative Hamiltonian constraint operator for LQC by only quantizing the Euclidean term \cite{Yang:2019ujs}. For the spatially flat Friedmann-Robertson-Walker (FRW) model with a massless scalar field, the backward evolution of the cosmological model determined by the effective Hamiltonian obtained in \cite{Yang:2019ujs} is bounced to an asymptotic de Sitter universe, and thus the classical big-bang singularity is resolved. However, the asymptotic de Sitter universe is associated with certain negative energy density of the matter field. One could therefore study whether the asymmetric bounce and the issue of negative energy density still holds if the Euclidean term and the Lorentzian term are treated independently in the new regularization scheme of LQC. This is the main motivation of the present paper. We will show that the new proposed Hamiltonian constraint operator, treating the Lorentzian term independently, can also drive an asymmetric quantum bounce evolution, and the backward evolution of the flat FRW cosmological model will be bounced to a de Sitter universe asymptotically with either positive or negative energy densities of matter depending on the values of the Barbero-Immirzi parameter.

The rest of this paper is organized as follows. In Sec. \ref{sec-II}, we briefly recall the elements of LQC. In Sec. \ref{sec-III}, we introduce a new alternative Hamiltonian constraint operator by treating the Euclidean term and the Lorentzian term independently in the new regularization scheme. Semiclassical analysis is implemented in Sec. \ref{sec-IV} to show that the new quantum dynamics has the correct classical limit, and the effective Hamiltonian constraint is obtained. In Sec. \ref{sec-V}, the effective Hamiltonian constraint is applied to study the effective dynamics of the LQC model. Our results are summarized and discussed in Sec. \ref{sec-VI}.

\section{Elements of loop quantum cosmology}
\label{sec-II}

For the spatially flat FRW model of LQC, one needs to choose an elementary cell ${\cal V}$ in order to avoid the noncompact structure of the spatial manifold, and to restrict all integrations to the cell. The volume of the cell measured by a fixed fiducial metric ${}^o\!q_{ab}$ is denoted by $V_o$. The classical variables, the connection $A^i_a$ and the density-weighted triad $\tilde{E}^a_i$, are reduced to \cite{Ashtekar:2003hd}
\begin{align}\label{eqn:c-p}
A^i_a=c V_o^{-\frac{1}{3}}\ {}^o\!\omega^i_a\,, \quad \tilde{E}^a_i=p V_o^{-\frac{2}{3}}\sqrt{{}^o\!q}\ {}^o\!e^a_i\,,
\end{align}
where $({}^o\!\omega^i_a,{}^o\!e^a_i)$ are a set of orthonormal cotriads and triads compatible with ${}^o\!q_{ab}$ and adapted to the edges of the elementary cell. The physical volume $V$ of the elemental cell ${\cal V}$ measured by the spatial metric $q_{ab}$ is related to $p$ via $V={|p|}^{3/2}$. The nontrivial Poisson bracket between the canonical variables $(c,p)$ reads
\begin{align}\label{eqn:c-p-bracket}
\{c,p\}=\frac{\kappa\gamma}{3}\,,
\end{align}
where $\gamma$ represents the Barbero-Immirzi parameter, and $\kappa=8\pi{G}$ with $G$ being the Newtonian's gravitational constant. In the improved scheme, it is convenient to choose the basic variables as \cite{Ding:2008tq}
\begin{align}
b&:=\frac{\bar\mu{c}}{2}\,, \qquad v:=\frac{{\rm sgn}(p)|p|^{3/2}}{2\pi|\gamma|\ell^2_{\rm p}\sqrt\Delta}\,,
\end{align}
where $\ell_{\rm p}\equiv\sqrt{G\hbar}$ is the Planck length, ${\rm sgn}(p)$ denotes the signature of $p$, $\Delta\equiv 4\sqrt{3}\,\pi|\gamma|\ell^2_{\rm p}$ is the area gap in full LQG \cite{Ashtekar:2008zu} \footnote{In the literature, the area gap $\Delta$ is often written as $4\sqrt{3}\,\pi\gamma\ell^2_{\rm p}$ under the assumption that the Barbero-Immirzi  parameter is positive. In the present paper, $\gamma$ takes any real value.}, $\bar\mu\equiv \sqrt{\Delta/|p|}$. Note that classically $b$ is related to the Hubble parameter $H$ by
\begin{align}\label{eq:b-H}
b=\frac{\gamma\sqrt{\Delta}}{2}H\,.
\end{align}
The Poisson bracket between $b$ and $v$ is given by
\begin{align}\label{eq:poisson-bracke}
\{b,v\}=\frac{{\rm sgn}(\gamma)}{\hbar}\,.
\end{align}
In the cosmological model, the only constraint which has to be satisfied is the Hamiltonian constraint. The gravitational Hamiltonian constraint reads
\begin{align}\label{eqn:full-hamilton}
{\cal H }_{\rm grav}&=\int_{\cal V}{\rm d}^3x\, \frac{\tilde{E}^a_i\tilde{E}^b_j}{2\kappa\sqrt{\textrm{det}(q)}}\left[{\epsilon^{ij}}_kF^{k}_{ab}
-2(1+\gamma^2)K^i_{[a}K^j_{b]}\right]\notag\\
&\equiv{\cal H}_{\rm grav}^{\rm E}-2(1+\gamma^2)\,{\cal H}_{\rm grav}^{\rm L}\,,
\end{align}
where $F^i_{ab}$ is the curvature of connection $A^i_a$, and $K^i_a$ represents the extrinsic curvature of the spatial manifold. The Euclidean term ${\cal H}_{\rm grav}^{\rm E}$ and the Lorentzian term ${\cal H}_{\rm grav}^{\rm L}$ of ${\cal H }_{\rm grav}$ can be expressed in terms of the reduced variables as 
\begin{align}\label{eq:Euclidean-term}
{\cal H}_{\rm grav}^{\rm E}&=\int_{\mathcal{V}}{\rm d}^3x\, \frac{\tilde{E}^a_i\tilde{E}^b_j}{2\kappa\sqrt{\textrm{det}(q)}}{\epsilon^{ij}}_kF^{k}_{ab}=\frac{3|\gamma|\hbar}{\sqrt{\Delta}}\,b^2|v|\,,
\end{align}
and
\begin{align}\label{eq:Lorentz-term}
{\cal H}_{\rm grav}^{\rm L}&=\int_{\mathcal{V}}{\rm d}^3x\, \frac{\tilde{E}^a_i\tilde{E}^b_j}{2\kappa\sqrt{\textrm{det}(q)}}K^i_{[a}K^j_{b]}=\frac{3\hbar}{2|\gamma|\sqrt{\Delta}}\,b^2|v|\,.
\end{align}
Hence, in the classical case, the two terms are proportional to each other.

Consider the FRW model with a massless scalar field. The Hamiltonian for the scalar field $T$ reads
\begin{align}\label{eqn:matter-hamilton}
{\cal H }_{\rm T}=\frac{p_T^2}{4\pi|\gamma|\ell_{\rm P}^2\sqrt{\Delta}\,|v|}\,,
\end{align}
where $p_T$ is the conjugate momentum of $T$ with the Poisson bracket $\{T,p_T\}=1$. Combining the gravity part with the matter part, the total Hamiltonian constraint is given by
\begin{align}\label{classical-Hamiltonian}
{\cal H}_{\rm tot}&=-\frac{3\hbar}{|\gamma|\sqrt{\Delta}}\,b^2|v|+\frac{p_T^2}{4\pi|\gamma|\ell_{\rm P}^2\sqrt{\Delta}\,|v|}\,.
\end{align}
In terms of Eq. \eqref{eq:b-H}, vanishing the Hamiltonian constraint yields the Friedmann equation 
\begin{align}
H^2=\frac{\kappa}{3}\frac{p_T^2}{2|p|^3}=:\frac{\kappa}{3}\,\rho_T\,.
\end{align}

To pass to the quantum theory, one first constructs the kinematical Hilbert space corresponding to the gravitational degrees of freedom, which is given by ${\mathscr H}_{\rm grav}=L^2(\mathbb{R}_{\mathrm{Bohr}},{\mathrm d}\mu_{\mathrm{Bohr}})$, where $\mathbb{R}_{\rm Bohr}$ is the Bohr compactification of the real line $\mathbb{R}$, and ${\mathrm d}\mu_{\mathrm{Bohr}}$ is the Haar measure \cite{Ashtekar:2003hd}. In ${\mathscr H}_{\rm grav}$, there are two elementary operators, $\widehat{e^{{\rm i}b}}$ and $\hat{v}$. In the $v$-representation, the actions of these two operators on the basis read
\begin{align}
\widehat{e^{{\rm i}b}}\,|v\rangle&=|v+1\rangle\,,\qquad\hat{v}\,|v\rangle=v\,|v\rangle\,.
\end{align}
The classical holonomy of the connection along an edge parallel to the vector ${}^o\!e^a_i$ with physical length $\sqrt{\Delta}$ reads
\begin{align}\label{eqn:i-holonomy}
&h^{(\bar\mu)}_i=\cos(b)\,\mathbb{I}+2\sin(b)\,\tau_i\,,
\end{align}
where $\mathbb{I}$ is the $2\times2$ identity matrix, and $\tau_i:=-\frac{{\rm i}}{2}\sigma_i$ with $\sigma_i$ being the Pauli matrices. In terms of $\widehat{e^{{\rm i}b}}$, one can easily to write down the action of the corresponding holonomy operator $\hat{h}^{(\bar\mu)}_i$ on $|v\rangle$. Then for the scalar filed, one can employ the Sch\"ordinger quantization. The scalar field $T$ and its conjugate momentum $p_T$ are quantized as a multiplication operator and a derivative operator on the Hilbert space $L^2(\mathbb{R},{\mathrm d} T)$  respectively as
\begin{align}
\hat{T}\cdot\psi(T)&=T\,\psi(T)\,,\qquad
\hat{p}_T\cdot\psi(T)=-{\rm i}\hbar\frac{{\mathrm d}}{{\mathrm d} T}\,\psi(T)\,,
\end{align}
where $\psi(T)\in L^2(\mathbb{R},{\mathrm d} T)$. Then the total kinematical Hilbert space is ${\mathscr H}_{\rm tot}={\mathscr H}_{\rm grav}\otimes L^2(\mathbb{R},{\mathrm d} T)$. 

\section{Alternative Hamiltonian constraint operator}
\label{sec-III}

Let us recall the regularization and quantization for the Euclidean term of the Hamiltonian constraint in \cite{Yang:2019ujs}. Following \cite{Soo:2005gw,Soo:2007hj}, by introducing the Chern-Simons functional
\begin{align}\label{csaction}
 S_{\rm cs}&=-\frac{1}{2}\int_{\cal V} {\rm d}^3x\, {\cal S}\,\tilde{\epsilon}^{abc}\left(F^i_{ab}A^i_c-\frac{1}{3}\epsilon_{ijk}A^i_aA^j_bA^k_c\right)\,,
\end{align}
the Euclidean term of Eq. \eqref{eq:Euclidean-term} can be reexpressed as
\begin{align}\label{eqn:cs-H(1)}
{\cal H}_{\rm grav}^{\rm E}=-\frac{1}{\kappa^2\gamma}\left\{S_{\rm cs},V\right\}\,,
\end{align}
where ${\cal S}\equiv{\rm sgn}\left[\det(e^i_a)\right]$, $\tilde{\epsilon}^{abc}$ denotes the Levi-Civita symbol. To regularize the expression \eqref{eqn:cs-H(1)}, we first express the Chern-Simons functional \eqref{csaction} in terms of holonomies as \cite{Yang:2019ujs}
\begin{align}\label{eqn:cs-discrete1}
S_{\rm cs}^{\rm reg}&=\frac{{\rm sgn}(p)}{\bar{\mu}^3}\left(8\sin^3(b)-6\sin^2{(2b)}\sin(b)\right)\,.
\end{align}
It should be noticed that in Eq. \eqref{eqn:cs-discrete1} we have taken into account the existence of the area gap $\Delta$, and thus chosen $\bar{\mu}^2|p|=\Delta$ \cite{Ashtekar:2006wn}. Hence, by Eq. \eqref{eqn:cs-H(1)} the regularized Euclidean Hamiltonian constraint reads
\begin{align}\label{eq:Euclidean-term-reg}
{\cal H }_{\rm grav}^{\rm E, reg}&=-\frac{\gamma\hbar^2}{16\sqrt{\Delta}}\left\{|v|{\rm sgn}(v)
\left(8\sin^3(b)\right.\right.\notag\\
&\hspace{2cm}\left.\left.-6\sin^2(2b)\sin(b)\right),|v|\right\}\,.
\end{align}
The corresponding Euclidean Hamiltonian constraint operator is obtained by replacing the classical variables with their quantum analogs as \cite{Yang:2019ujs}
\begin{align}\label{eqn:H(1)-operator}
\hat{\cal H}_{\rm grav}^{\rm E}=\frac{{\rm i}\gamma\hbar}{32\sqrt{\Delta}}\,\hat{|v|}^{1/2}
\left[\hat{\cal B}\,{\rm sgn}(\hat{v})+{\rm sgn}(\hat{v})\,\hat{\cal B},\hat{|v|}\right]\hat{|v|}^{1/2}\,,
\end{align}
where
\begin{align}
\hat{\cal B}\equiv 8\widehat{\sin^3(b)}-6\widehat{\sin(2b)}\;\widehat{\sin(b)}\;\widehat{\sin(2b)}\,.
\end{align}
Its action on the state $|v\rangle$ is given by
\begin{align}\label{eq:Euclidean-action}
\hat{\cal H}_{\rm grav}^{\rm E}\,|v\rangle=\sum_{k\in\{-5,-3,\cdots,5\}}\lambda_{|k|}f_k(v)\,|v+k\rangle\,,
\end{align}
where $\lambda_5=3, \lambda_3=-7, \lambda_1=6$, and
\begin{align}
f_k(v)&:=\frac{\gamma\hbar}{128\sqrt{\Delta}}{\rm sgn}(k)\left[{\rm sgn}(v)+{\rm sgn}(v+ k)\right]M_{0,k}(v)\notag\\
&\hspace{1.5cm}\times \sqrt{|v(v+ k)|}\,,
\end{align} 
with
\begin{align}
M_{a,b}(v):=|v+a|-|v+b|\,.
\end{align}

Now we turn to the Lorentzian term ${\cal H}_{\rm grav}^{\rm L}$ of Eq. \eqref{eq:Lorentz-term}. Since ${\cal H}_{\rm grav}^{\rm L}$ is not proportional to the Euclidean term ${\cal H}_{\rm grav}^{\rm E}$ in the full theory, it is reasonable to treat ${\cal H}_{\rm grav}^{\rm L}$ differently from ${\cal H}_{\rm grav}^{\rm E}$ by expressing the extrinsic curvature $K^i_a$ in Eq. \eqref{eq:Lorentz-term} in a form similar to that of full LQG. Thus the Lorentzian term \eqref{eq:Lorentz-term} can be reexpressed as
\begin{align}\label{eq:Lorentz-term-exp-B}
{\cal H}_{\rm grav}^{\rm L}&=\int_{\mathcal{V}}{\rm d}^3x\, \frac{\tilde{E}^a_i\tilde{E}^b_j}{2\kappa\sqrt{\textrm{det}(q)}}K^i_{[a}K^j_{b]}\notag\\
&=\frac{3}{2\kappa^3\gamma^6}|p|^{1/2}\{c,B\}\{c,B\}\notag\\
&=-\frac{1}{\kappa^3\gamma^6}|p|^{1/2}\sum_{i=1}^3{\rm Tr}\left(\{c\tau_i,B\}\{c\tau_i,B\}\right)\notag\\
&=-\frac{4 |p|^{1/2}}{9\kappa^3\gamma^6\bar{\mu}^2}\sum_{i=1}^3{\rm Tr}\left(h^{(\bar\mu)}_i\{{h^{(\bar\mu)}_i}^{-1},B\}h^{(\bar\mu)}_i\{{h^{(\bar\mu)}_i}^{-1},B\}\right)\notag\\
&=-\frac{4 V}{9\kappa^3\gamma^6\Delta}\sum_{i=1}^3{\rm Tr}\left(h^{(\bar\mu)}_i\left\{{h^{(\bar\mu)}_i}^{-1},\{{\cal H}_{\rm grav}^{\rm E},V\}\right\}\right.\notag\\
&\hspace{3cm}\times\left.h^{(\bar\mu)}_i\left\{{h^{(\bar\mu)}_i}^{-1},\{{\cal H}_{\rm grav}^{\rm E},V\}\right\}\right)\,,
\end{align}
where we defined $B:=\gamma^2\int_{\cal V}{\rm d}^3x\,K^i_a\tilde{E}^a_i=3\gamma cp$ and used \cite{Thiemann:1996aw}
\begin{align}
 B=\{{\cal H}_{\rm grav}^{\rm E},V\}\,,
\end{align}
as well as \cite{Yang:2009fp}
\begin{align}
\{c\tau_i,B\}=-\frac{2}{3\bar{\mu}}h^{(\bar\mu)}_i\{{h^{(\bar\mu)}_i}^{-1},B\}\,.
\end{align}
Here the holonomy $h^{(\bar\mu)}_i$ should be understood as a function of both $b$ and $v$. Replacing ${\cal H}_{\rm grav}^{\rm E}$ in Eq. \eqref{eq:Lorentz-term-exp-B} by its regularized version \eqref{eq:Euclidean-term-reg}, we obtain the regularized version of ${\cal H}_{\rm grav}^{\rm L}$ as
\begin{align}\label{eq:Lorentz-term-reg}
{\cal H}_{\rm grav}^{\rm L, reg}
&=-\frac{\hbar^3\sqrt{\Delta}}{144|\gamma|\gamma^2}\,|v|\,\sum_{i=1}^3{\rm Tr}\left(h^{(\bar\mu)}_i\left\{{h^{(\bar\mu)}_i}^{-1},\{{\cal H}_{\rm grav}^{\rm E, reg},|v|\}\right\}\right.\notag\\
&\hspace{3cm}\times\left.h^{(\bar\mu)}_i\left\{{h^{(\bar\mu)}_i}^{-1},\{{\cal H}_{\rm grav}^{\rm E, reg},|v|\}\right\}\right)\,.\notag\\
\end{align}
Then we can directly write down the corresponding quantum operator for the Lorentzian term as
\begin{align}
\hat{\cal H}_{\rm grav}^{\rm L}&=-\frac{\sqrt{\Delta}}{144|\gamma|\gamma^2\hbar}\,\sum_{i=1}^3{\rm Tr}\left(\widehat{h^{(\bar\mu)}_i}\left[\widehat{{h^{(\bar\mu)}_i}^{-1}},[\hat{\cal H}_{\rm grav}^{\rm E},|\hat{v}|]\right]|\hat{v}|\right.\notag\\
&\hspace{3cm}\times\left.\widehat{h^{(\bar\mu)}_i}\left[\widehat{{h^{(\bar\mu)}_i}^{-1}},[\hat{\cal H}_{\rm grav}^{\rm E},|\hat{v}|]\right]\right)\notag\\
&=\frac{\sqrt{\Delta}}{24|\gamma|\gamma^2\hbar}\left(\hat{O}_{[\hat{\cal H}_{\rm grav}^{\rm E},|\hat{v|}]}\,\hat{|v|}\,\hat{O}_{[\hat{\cal H}_{\rm grav}^{\rm E},|\hat{v|}]}\right.\notag\\
&\hspace{2cm}\left.-\hat{Q}_{[\hat{\cal H}_{\rm grav}^{\rm E},|\hat{v|}]}\,\hat{|v|}\,\hat{Q}_{[\hat{\cal H}_{\rm grav}^{\rm E},|\hat{v|}]}\right)\,,
\end{align}
where
\begin{align}
\hat{O}_{[\hat{\cal H}_{\rm grav}^{\rm E},|\hat{v|}]}&:=\widehat{\sin(b)}\,[\hat{\cal H}_{\rm grav}^{\rm E},|\hat{v|}]\,\widehat{\cos{b}}\notag\\
&\hspace{1cm}-\widehat{\cos(b)}\,[\hat{\cal H}_{\rm grav}^{\rm E},|\hat{v|}]\,\widehat{\sin(b)}\,,\\
\hat{Q}_{[\hat{\cal H}_{\rm grav}^{\rm E},|\hat{v|}]}&:=[\hat{\cal H}_{\rm grav}^{\rm E},|\hat{v|}]-\widehat{\sin(b)}\,[\hat{\cal H}_{\rm grav}^{\rm E},|\hat{v|}]\,\widehat{\sin{b}}\notag\\
&\hspace{1cm}-\widehat{\cos(b)}\,[\hat{\cal H}_{\rm grav}^{\rm E},|\hat{v|}]\,\widehat{\cos(b)}\,.
\end{align}
The actions of the operators $\hat{O}_{[\hat{\cal H}_{\rm grav}^{\rm E},|\hat{v|}]}$ and $\hat{Q}_{[\hat{\cal H}_{\rm grav}^{\rm E},|\hat{v|}]}$ on the basis $|v\rangle$ read
\begin{align}
\hat{O}_{[\hat{\cal H}_{\rm grav}^{\rm E},|\hat{v|}]}\, |v\rangle&=\frac{1}{2\,{\rm i}}\sum_{k\in\{-5,-3,\cdots,3,5\}}\lambda_{|k|}F^1_k(v)\,|v+k\rangle\,,\\
\hat{Q}_{[\hat{\cal H}_{\rm grav}^{\rm E},|\hat{v|}]\, |v\rangle}&=-\frac{1}{2}\sum_{k\in\{-5,-3,\cdots,3,5\}}\lambda_{|k|}F^2_k(v)\,|v+k\rangle\,,
\end{align}
where 
\begin{align}
F^1_k(v)&:=M_{0,k}(v-1)f_k(v-1)-M_{0,k}(v+1)f_k(v+1)\,,\\
F^2_k(v)&:=M_{0,k}(v-1)f_k(v-1)-2M_{0,k}(v)f_k(v)\notag\\
&\hspace{0.3cm}+M_{0,k}(v+1)f_k(v+1)\,.
\end{align}
Therefore, the action of the operator $\hat{\cal H}_{\rm grav}^{\rm L}$ on the basis $|v\rangle$ is given by
\begin{align}\label{eq:Lorentz-action}
\hat{\cal H}_{\rm grav}^{\rm L}\, |v\rangle&=-\frac{\sqrt{\Delta}}{96|\gamma|\gamma^2\hbar}\sum_{i=1,2}\sum_{k\in\{-5,-3,\cdots,3,5\}}\lambda_{|k|}F^i_k(v)\,|v+k|\notag\\
&\hspace{1.4cm}\times\sum_{l\in\{-5,-3,\cdots,3,5\}}\lambda_{|l|}F^i_l(v+k)\,|v+k+l\rangle\notag\\
&=:\sum_{i=1,2}\sum_{m\in\{-10,-8,\cdots,8,10\}}\tilde{F}^i_m(v)\,|v+m\rangle\,,
\end{align}
where
\begin{align}
\tilde{F}^i_{-10}(v)&=-\frac{3 \sqrt{\Delta }}{32 |\gamma|\gamma^2 \hbar} F^i_{-5}(v-5) F^i_{-5}(v) \left| v-5\right|\,,\notag\\
\tilde{F}^i_{-8}(v)&=\frac{7 \sqrt{\Delta }}{32 |\gamma|\gamma^2 \hbar } \left[F^i_{-5}(v) F^i_{-3}(v-5) \left| v-5\right|\right.\notag\\
                         &\hspace{1.65cm}\left.+F^i_{-5}(v-3) F^i_{-3}(v)\left| v-3\right|\right]\,,\notag\\
\tilde{F}^i_{-6}(v)&=-\frac{\sqrt{\Delta }}{96 |\gamma|\gamma^2 \hbar } \left[49 F^i_{-3}(v-3) F^i_{-3}(v) \left| v-3\right|\right.\notag\\
                         &\hspace{1.55cm}+18 F^i_{-5}(v) F^i_{-1}(v-5) \left| v-5\right|\notag\\ 
                         &\hspace{1.55cm}\left.+18 F^i_{-5}(v-1) F^i_{-1}(v) \left| v-1\right|\right]\,,\notag\\
\tilde{F}^i_{-4}(v)&=\frac{\sqrt{\Delta }}{16 |\gamma|\gamma^2 \hbar } \left[7 F^i_{-3}(v) F^i_{-1}(v-3) \left| v-3\right|\right.\notag\\
                         &\hspace{1.55cm}+7 F^i_{-3}(v-1) F^i_{-1}(v) \left| v-1\right|\notag\\ 
                         &\hspace{1.55cm}-3 F^i_{-5}(v) F^i_1(v-5) \left| v-5\right|\notag\\ 
                         &\hspace{1.55cm}\left.-3 F^i_{-5}(v+1) F^i_1(v) \left| v+1\right|\right]\,,\notag\\
\tilde{F}^i_{-2}(v)&=\frac{\sqrt{\Delta }}{32 |\gamma|\gamma^2 \hbar } \left[-12 F^i_{-1}(v-1) F^i_{-1}(v) \left| v-1\right|\right.\notag\\
                         &\hspace{1.55cm}+14 F^i_{-3}(v) F^i_1(v-3) \left| v-3\right|\notag\\ 
                         &\hspace{1.55cm}+14 F^i_{-3}(v+1) F^i_1(v) \left| v+1\right|\notag\\ 
                         &\hspace{1.55cm}+7 F^i_{-5}(v) F^i_3(v-5) \left| v-5\right|\notag\\ 
                         &\hspace{1.55cm}\left.+7 F^i_{-5}(v+3) F^i_3(v) \left| v+3\right|\right]\,,\notag\\
\tilde{F}^i_{0}(v)&=-\frac{\sqrt{\Delta }}{96 |\gamma|\gamma^2 \hbar } \left[36 F^i_{-1}(v) F^i_1(v-1) \left| v-1\right|\right.\notag\\
                         &\hspace{1.55cm}+36 F^i_{-1}(v+1) F^i_1(v) \left| v+1\right|\notag\\ 
                         &\hspace{1.55cm}+49 F^i_{-3}(v) F^i_3(v-3) \left| v-3\right|\notag\\
                         &\hspace{1.55cm}+49 F^i_{-3}(v+3) F^i_3(v) \left| v+3\right|\notag\\
                         &\hspace{1.55cm}+9 F^i_{-5}(v) F^i_5(v-5) \left| v-5\right|\notag\\ 
                         &\hspace{1.55cm}\left.+9 F^i_{-5}(v+5) F^i_5(v) \left| v+5\right|\right]\,,\notag\\
\tilde{F}^i_{2}(v)&=\frac{ \sqrt{\Delta }}{32 |\gamma|\gamma^2 \hbar } \left[-12 F^i_1(v) F^i_1(v+1) \left| v+1\right|\right.\notag\\
                         &\hspace{1.55cm}+14 F^i_{-1}(v) F^i_3(v-1) \left| v-1\right|\notag\\ 
                         &\hspace{1.55cm}+14 F^i_{-1}(v+3) F^i_3(v) \left| v+3\right|\notag\\ 
                         &\hspace{1.55cm}+7 F^i_{-3}(v) F^i_5(v-3) \left| v-3\right|\notag\\ 
                         &\hspace{1.55cm}\left.+7 F^i_{-3}(v+5) F^i_5(v) \left| v+5\right|\right]\,,\notag\\
\tilde{F}^i_{4}(v)&=\frac{\sqrt{\Delta }}{16 |\gamma|\gamma^2 \hbar } \left[7 F^i_1(v+3) F^i_3(v) \left| v+3\right|\right.\notag\\
                         &\hspace{1.55cm}+7 F^i_1(v) F^i_3(v+1) \left| v+1\right|\notag\\ 
                         &\hspace{1.55cm}-3 F^i_{-1}(v) F^i_5(v-1) \left| v-1\right|\notag\\ 
                         &\hspace{1.55cm}\left.-3 F^i_{-1}(v+5) F^i_5(v) \left| v+5\right|\right]\,,\notag\\
\tilde{F}^i_{6}(v)&=-\frac{\sqrt{\Delta }}{96 |\gamma|\gamma^2 \hbar } \left[49 F^i_3(v) F^i_3(v+3) \left| v+3\right|\right.\notag\\
                         &\hspace{1.55cm}+18 F^i_1(v+5) F^i_5(v) \left| v+5\right|\notag\\ 
                         &\hspace{1.55cm}\left.+18 F^i_1(v) F^i_5(v+1) \left| v+1\right|\right]\,,\notag\\
\tilde{F}^i_{8}(v)&=\frac{7 \sqrt{\Delta }}{32 |\gamma|\gamma^2 \hbar } \left[F^i_3(v+5) F^i_5(v) \left| v+5\right|\right.\notag\\
                         &\hspace{1.65cm}\left.+F^i_3(v) F^i_5(v+3) \left| v+3\right|\right]\,,\notag\\
\tilde{F}^i_{10}&=-\frac{3  \sqrt{\Delta }}{32 |\gamma|\gamma^2 \hbar}F^i_5(v) F^i_5(v+5) \left| v+5\right|\,.
\end{align}
For the convenience of semiclassical analysis, we also present the relations among the above coefficients as follows:
\begin{align}
M_{0,|k|}(v)&=-M_{0,-|k|}(v+|k|)\,,\\
f_{|k|}(v)&=f_{-|k|}(v+|k|)\,,\\
F^i_{|k|}(v)&=-F^i_{-|k|}(v+|k|)\,,\\
\tilde{F}^i_{|k|}(v)&=\tilde{F}^i_{-|k|}(v+|k|)\,.
\end{align}

Notice that the inverse volume operator corresponding to $1/V$ appearing in the Hamiltonian \eqref{eqn:matter-hamilton} of the scalar field is given by \cite{Ashtekar:2006wn}
\begin{align}\label{eq:inverseV}
\widehat{V^{-1}}\,|v\rangle=\frac{C(v)}{2\pi |\gamma|\sqrt{\Delta}\ell_{\mathrm p}^2}\,|v\rangle\,,
\end{align}
where
\begin{align}
C(v)\equiv \left(\frac{3}{2}\right)^3|v|
\left|{|v+1|^{{1}/{3}}}-{|v-1|^{{1}/{3}}}\right|^3\,.
\end{align}
Then the action of the Hamiltonian constraint operator of the matter part on a quantum state $|\psi\rangle=\psi(v;T)\,|v\rangle\otimes|T\rangle$ reads
\begin{align}\label{hamilton_matter_op}
\hat{\cal H}_{\rm M}\cdot\psi(v;T)=-\frac{\hbar^2}{4\pi|\gamma|\sqrt{\Delta}\ell_{\mathrm p}^2}C(v)\partial^2_T\psi(v;T)\,.
\end{align}
Finally, by using Eqs. \eqref {eq:Euclidean-action}, \eqref {eq:Lorentz-action}, and \eqref{hamilton_matter_op}, we have the total Hamiltonian constraint equation
\begin{align}\label{eq:quantum-equation}
 \hat{\cal H}_{\rm tot}\cdot\psi(v;T)\equiv\left(\hat{\cal H}_{\rm grav}+\hat{\cal H}_{\rm M}\right)\cdot\psi(v;T)=0\,,
\end{align}
which gives the quantum dynamics of the coupled system.

\section{Semiclassical analysis of the quantum dynamics}
\label{sec-IV}

In the previous section, we have quantized the Hamiltonian constraint and obtained the quantum dynamics, which is determined by the difference equation \eqref{eq:quantum-equation}. In this section, we will show that the quantum dynamics has the correct classical limit by semiclassical analysis, and obtain the effective Hamiltonian constraint. To this aim, we will calculate the expectation value of $\hat{\cal H}_{\rm tot}$ under certain coherent state \cite{Ashtekar:2002sn,Ashtekar:2003hd,Ding:2008tq}.

A coherent state peaked at a point $(b_o,v_o,T_o,p_T)$ in the classical phase space reads \cite{Ding:2008tq}
\begin{align}\label{eqn:dual-semiclassical-state}
\left(\Psi_{(b_o,v_o,T_o,p_T)}\right|&:=\int {\rm d}T\sum_{v\in\mathbb{R}}e^{-\frac{\epsilon^2}{2}(v-v_o)^2} e^{{\rm i}b_o(v-v_o)}\notag\\
&\hspace{1cm}\times  e^{-\frac{\sigma^2}{2}(T-T_o)^2}e^{\frac{\rm i}{\hbar}p_T(T-T_o)}(v|\otimes (T|\,,
\end{align}
where $\epsilon$ and $\sigma$ are the Gaussian spreads in the gravitational sector and scalar field sector, respectively. For practical calculations, one only needs to use the shadow of the state on the regular lattice with spacing 1 as
\begin{align}\label{eqn:shadow-grav}
\left|\Psi\right\rangle&:=\int{\rm d}T\sum_{n\in\mathbb{Z}}e^{-\frac{\epsilon^2}{2}(n-v_o)^2}e^{-{\rm i}b_o(n-v_o)}\notag\\
&\hspace{1cm}\times e^{-\frac{\sigma^2}{2}(T-T_o)^2} e^{-\frac{\rm i}{\hbar}p_T(T-T_o)}\,|n\rangle\otimes |T\rangle\,.
\end{align}
The parameters in the coherent state \eqref{eqn:dual-semiclassical-state} are restricted to satisfying $v_o\gg 1$, $\epsilon\ll b_o\ll 1$, $v_o\epsilon\gg 1$, $\sigma\ll T_o$ and $p_T\sigma\gg 1$ in order to make the state be sharply peaked in the classical phase space of the universe with large volume.

\begin{figure}
  \includegraphics[width=\columnwidth]{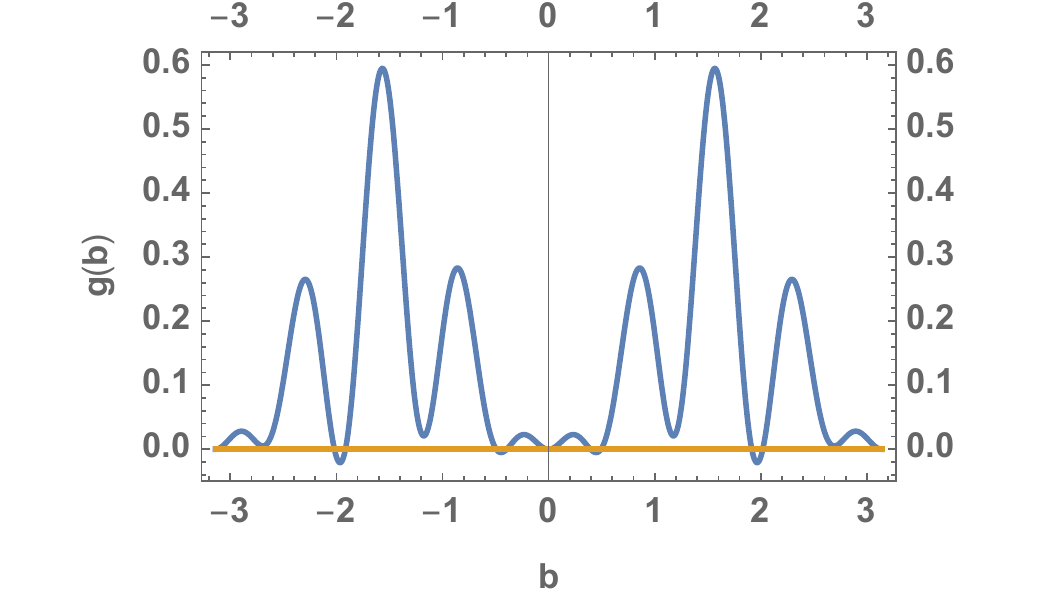}
  \caption{Plot of  $g(b)$ for $\gamma=0.2375$ with $b\in [-\pi,\pi]$. $g(b)=0$ yields $b\in\{0,\pm 0.394457,\pm 0.493439,\pm 1.91495,\pm 2.01833,\pm\pi\}$ .}
  \label{fig:g-v}
\end{figure}

Now let us calculate the expectation value of $\hat{\cal H}_{\rm grav}$. Recall that the expectation value of $\hat{\cal H }_{\rm grav}^{\rm E}$ was calculated in \cite{Yang:2019ujs} as
\begin{align}\label{eq:expection-Euclidean}
{\cal H}^{\rm E, eff}_{\rm grav}&:=\frac{\langle \Psi|\hat{\cal H }_{\rm grav}^{\rm E}|\Psi\rangle}{\langle \Psi|\Psi\rangle}\notag\\
&= \frac{3|\gamma|\hbar}{8\sqrt{\Delta}}v_o\left[3\cos(b_o)+5\cos(3b_o)\right]\sin^2(b_o).
\end{align}
Thus we only need to deal with the Lorentzian term $\hat{\cal H}_{\rm grav}^{\rm L}$. From Eq. \eqref{eq:Lorentz-action}, one can easily write down the action of $\hat{\cal H}_{\rm grav}^{\rm L}$ on the shadow state as
\begin{align}
&\hat{\cal H}_{\rm grav}^{\rm L}|\Psi\rangle\notag\\
&=\sum_{n\in\mathbb{Z}}e^{-\frac{\epsilon^2}{2}(n-v_o)^2}e^{-{\rm i}b_o(n-v_o)}\notag\\
&\quad\times\sum_{i=1,2}\Big[\sum_{k\in\{2,4,\cdots,10\}}\left( \tilde{F}^i_k(n)
\,|n+ k\rangle+ \tilde{F}^i_{-k}(n) \,|n-k\rangle\right)\notag\\
&\hspace{1cm}+ \tilde{F}^i_0(n) \,|n\rangle\Big]\,.
\end{align}
By the relation $\tilde{F}^i_{|k|}(v)=\tilde{F}^i_{-|k|}(v+|k|)$, we then obtain
\begin{align}
(\Psi|\hat{\cal H }_{\rm grav}^{\rm L}|\Psi\rangle&=2\sum\limits_{k\in\{2,4,\cdots,10\}}e^{-\frac{\epsilon^2}{4}k^2}\cos{(kb_o)}\notag\\
&\hspace{2cm}\times\sum_{i=1,2}\sum\limits_{n\in\mathbb{Z}}e^{-\epsilon^2\left(n-v_o+\frac{k}{2}\right)^2} \tilde{F}^i_k(n)\notag\\
&\quad+\sum_{i=1,2}\sum\limits_{n\in\mathbb{Z}}e^{-\epsilon^2\left(n-v_o\right)^2} \tilde{F}^i_0(n)\notag\\
&=:2\sum\limits_{k\in\{2,4,\cdots,10\}}e^{-\frac{\epsilon^2}{4}k^2}\cos{(kb_o)}I_k+I_0\,.
\end{align}
Applying the Poisson resummation formula and the steepest decent method, we get the following 
approximative values of the factors $I_l$ to leading order as,
\begin{align}
I_0&\approx\frac{\sqrt{\pi}}{\epsilon}\frac{1605 \hbar v_o}{8192\, |\gamma| \sqrt{\Delta }}\,,\notag\\
I_2&\approx-\frac{\sqrt{\pi}}{\epsilon}\frac{1707 \hbar v_o }{16384\, |\gamma| \sqrt{\Delta }}\,,\notag\\
I_4&\approx\frac{\sqrt{\pi}}{\epsilon}\frac{69 \hbar v_o }{4096\, |\gamma| \sqrt{\Delta }}\,,\notag\\
I_6&\approx-\frac{\sqrt{\pi}}{\epsilon}\frac{1623 \hbar v_o }{32768\, |\gamma| \sqrt{\Delta }}\,,\notag\\
I_8&\approx\frac{\sqrt{\pi}}{\epsilon}\frac{1575 \hbar v_o }{16384\, |\gamma| \sqrt{\Delta }}\,,\notag\\
I_{10}&\approx-\frac{\sqrt{\pi}}{\epsilon}\frac{1875 \hbar v_o }{32768\, |\gamma| \sqrt{\Delta }}\,.
\end{align}
Thus, we have
\begin{align}\label{eq:expection-Lorentz}
{\cal H }_{\rm grav}^{\rm L, eff}&:=\frac{\langle \Psi|\hat{\cal H }_{\rm grav}^{\rm L}|\Psi\rangle}{\langle \Psi|\Psi\rangle}\notag\\
&\approx-\frac{3\hbar v_o}{16384\,|\gamma| \sqrt{\Delta}}\left[1138 \cos \left(2 b_o\right)-184 \cos \left(4 b_o\right)\right.\notag\\
&\hspace{2.5cm}+541 \cos \left(6 b_o\right)-1050 \cos \left(8 b_o\right)\notag\\
&\hspace{2.5cm}\left.+625 \cos \left(10 b_o\right)-1070\right]\notag\\
&= \frac{3  \hbar v_o} {8192\, |\gamma| \sqrt{\Delta }} \left[2 \sin \left(b_o\right)-21 \sin \left(3 b_o\right)+25 \sin \left(5 b_o\right)\right]^2.
\end{align}
It is worth noting that the Gaussian variances $\epsilon$ and $\sigma$ do not occur in the leading-order correction in \eqref{eq:expection-Lorentz},  which is concerned in the current work, although they would appear in the higher-order corrections. Hence, by dropping the subscript $o$ for simplicity, the effective Hamiltonian constraint of gravitational part reads
\begin{align}\label{eq:expection-gravity}
{\cal H}_{\rm grav}^{\rm Full, eff}=&{\cal H }_{\rm grav}^{\rm E, eff}-2(1+\gamma^2){\cal H }_{\rm grav}^{\rm L, eff}\notag\\
=&\frac{3|\gamma|\hbar}{8\sqrt{\Delta}}\,v\left[3\cos(b)+5\cos(3b)\right]\sin^2(b)\notag\\
&-\frac{3(1+\gamma^2)\hbar}{4096\,|\gamma| \sqrt{\Delta}}\,v\left[2\sin(b)-21\sin(3b)+25\sin(5b)\right]^2\notag\\
=&-\frac{3\hbar v}{|\gamma|\sqrt{\Delta}}g(b)\,,
\end{align}
where
\begin{align}\label{eq:g-b}
g(b)&:=\frac{\sin^2(b)}{2048}\left\{-256 \gamma^2 \left[3 \cos (b)+5 \cos (3 b)\right]\right.\notag\\
&\hspace{0.7cm}\left.+2 (1+\gamma^2) \left[4 \cos (2 b)+25 \cos (4 b)+3\right]^2\right\}\,.
\end{align}
The function $g(b)$ is plotted in \figref{fig:g-v}.  Expanding $g(b)$ at $b=0$ yields
\begin{align}\label{eq:gb-b}
g(b)=b^2+O(b^3)\,,
\end{align}
which implies that $g(b)$ contains the quantum correction in addition to $b^2$. 

Taking into account the result for the matter sector given in \citep{Ashtekar:2002sn,Ding:2008tq}, the resulting expectation value of the total Hamiltonian constraint reads
\begin{align}\label{eq:Full-eff-Hamiltonian}
{\cal H}^{\rm Full,  eff}_{\rm tot}=&{\cal H}_{\rm grav}^{\rm Full, eff}+{\cal H }_{\rm T}=-\frac{3\hbar v}{|\gamma|\sqrt{\Delta}}g(b)+\frac{p_T^2}{4\pi|\gamma|G\hbar\sqrt{\Delta}\,v}\,.
\end{align}
In terms of Eq. \eqref{eq:gb-b}, the effective Hamiltonian constraint ${\cal H}^{\rm Full,  eff}_{\rm tot}$ in Eq. \eqref{eq:Full-eff-Hamiltonian} reduces to its classical expression ${\cal H}_{\rm tot}$ of Eq. \eqref{classical-Hamiltonian} as $b\rightarrow0$. Hence the quantum dynamics has the correct classical limit. Note that it is straightforward to check that the effective Hamiltonians \eqref{eq:expection-Lorentz} and \eqref{eq:expection-gravity} coincide with the regularized expressions \eqref{eq:Euclidean-term-reg} and \eqref{eq:Lorentz-term-reg} respectively.

\section{Dynamical analysis}
\label{sec-V}

\begin{figure}
  \includegraphics[width=\columnwidth]{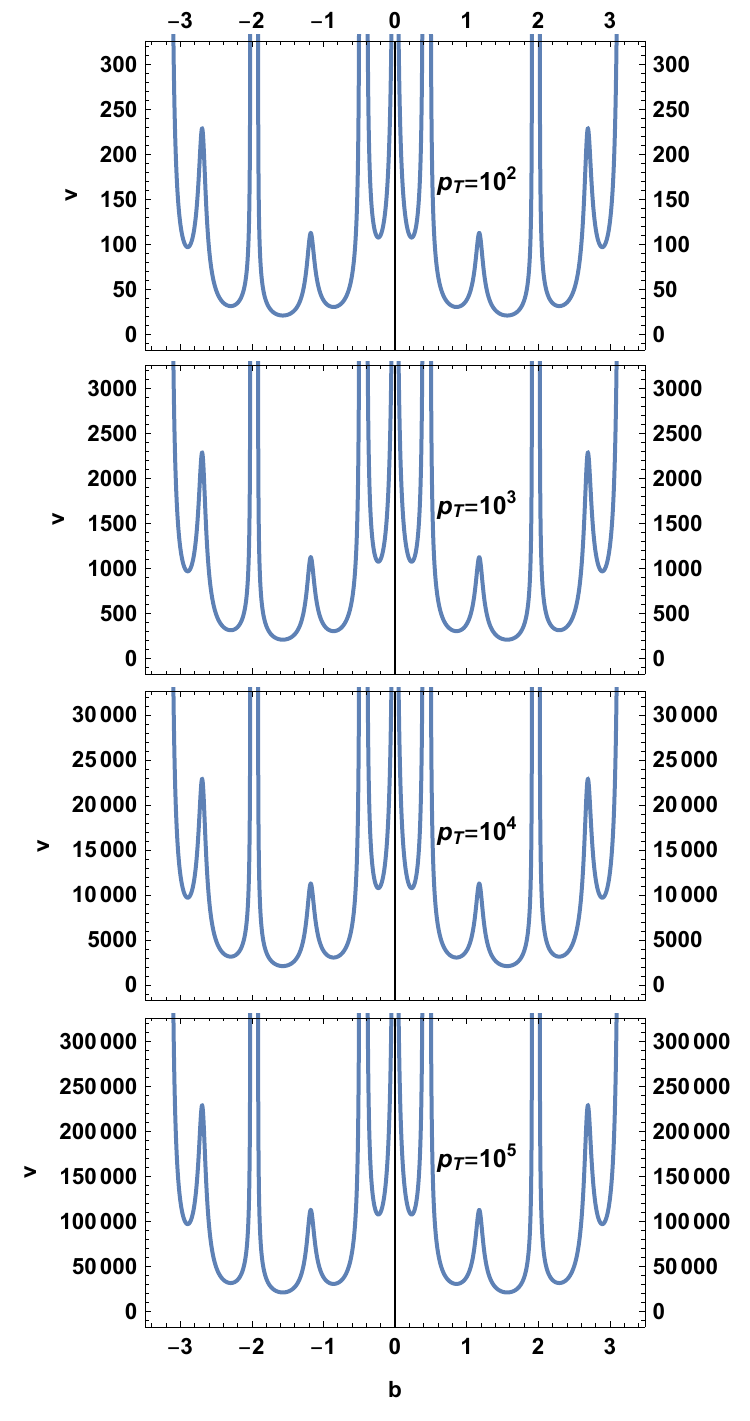}
  \caption{Plots of the constraint equation ${\cal H}^{\rm Full, eff}_{\rm tot}=0$ for different values of $p_T$, where $G=\hbar=1$, $\gamma=0.2375$. It is shown that $v=0$ cannot be a solution.}
  \label{fig:b-v}
\end{figure}

From the effective Hamiltonian constraint \eqref{eq:Full-eff-Hamiltonian}, it is easy to see that $p_T$ is a constant of motion. Thus, $T$ is a monotonic function of the cosmological time. Hence the matter field $T$ can be regarded as an internal clock, by which the relative evolution can be defined. In addition, it is obvious from Eq. \eqref{eq:Full-eff-Hamiltonian} that $v=0$ can never be a solution to the constraint equation
\begin{align}\label{eq:constraint-eq}
{\cal H}^{\rm Full, eff}_{\rm tot}=0\,.
\end{align}
This indicates that the classical big-bang singularity at $v=0$ will be resolved by Eq. \eqref{eq:constraint-eq}. For some given $p_T$, the evolutions of $v$ with respect to $b$ determined by the effective Hamiltonian constraint are plotted in \figref{fig:b-v}, in which the conclusion of $v\neq 0$ is confirmed.

By Eq. \eqref{eq:constraint-eq}, the matter density can be expressed as
\begin{align}\label{rho-relation}
 \rho_T(v)&=\frac{p_T^2}{2V^2}=\frac{3}{2 \pi G \gamma^2 \Delta}g(b)=:\rho^{\rm Full, eff}_T(b)\,.
\end{align}
In the following, we focus on the region $b\in[0,\pi]$, in which our universe lives. Equation \eqref{rho-relation} requires that $g(b)\geqslant0$. Hence, Fig. \ref{fig:g-v} indicates that we can divide the region $[0,\pi]$ of $b$ into three unconnected regions, depending only on the parameter $\gamma$. In the case of $\gamma=0.2375$, the three regions read
\begin{align}
\text{Region I:}&\qquad [0, 0.394457]\,,\notag\\
\text{Region II:}&\qquad [0.493439, 1.91495]\,,\notag\\
\text{Region III:}&\qquad [2.01833, \pi]\,.\notag
\end{align}
In the three regions, the local maximal values of $\rho^{\rm Full, eff}_T(b)$, or the critical matter densities $\rho_{\rm crit}$, read respectively
\begin{align}
\text{Region I:}&\qquad \rho^{\rm I}_{T,\rm crit}=\frac{4.85899}{8 \pi  G \Delta},\quad \text{at}\;b=0.23222\,,\notag\\
\text{Region II:}&\qquad \rho^{\rm II,1}_{T,\rm crit}=\frac{60.174}{8 \pi  G \Delta},\quad \text{at}\; b=0.858344\,,\notag\\
                        &\qquad \rho^{\rm II,2}_{T,\rm crit}=\frac{126.455}{8 \pi  G \Delta},\quad \text{at}\; b=1.56661\,,\notag\\
\text{Region III:}&\qquad \rho^{\rm III,1}_{T,\rm crit}=\frac{56.4203}{8 \pi  G \Delta},\quad \text{at}\; b=2.29351\,,\notag\\
                          &\qquad \rho^{\rm III,2}_{T,\rm crit}=\frac{5.98796}{8 \pi  G \Delta},\quad \text{at}\; b=2.89445\,.\notag
\end{align}
The current value of the Hubble parameter from the observation is $H_0\approx2.18553\times10^{-18}\,{\rm s}^{-1}$ \cite{Aghanim:2018eyx}, which indicates that the universe where we live locates in the region I.

Now let us study the asymptotic behavior of the effective dynamics in the region I at the large $v$ limit. For $v\rightarrow \infty$, the matter density $\rho_T(v)$ in Eq. \eqref{rho-relation} goes to zero, and thus
\begin{align}
{\cal H}_{\rm grav}^{\rm Full, eff} \;\rightarrow\;0\,\qquad\Leftrightarrow\qquad g(b)\;\rightarrow\;0\,.
\end{align}
This leads to
\begin{align}\label{eqn:bb}
b\;\rightarrow\;b_0=
  \left\{
  \begin{array}{ll}
  b_{\rm c,I,1}\equiv0\\
  b_{\rm c,I,2}\equiv0.394457
 \end{array}\right.\,.
\end{align}
Hence there are two types of classical universe, namely the type-I-1 universe and the type-I-2 universe. Expanding ${\cal H}_{\rm tot}^{\rm Full, eff}$ at $b_0$ up to the second order yields the classical behavior of the effective Hamiltonian constraint as
\begin{align}
{\cal H}_{\rm tot}^{\rm Full, eff}\;\rightarrow\;&-\frac{3\hbar v}{|\gamma|\sqrt{\Delta}}\left[g'(b_0)(b-b_0)+\frac{g''(b_0)}{2}(b-b_0)^2\right]\notag\\
&+\frac{p_T^2}{4\pi|\gamma|G\hbar\sqrt{\Delta}\,v}\,,
\end{align}
where $'$ and $''$ denote the first-order and second-order derivatives of $g(b)$ with respective to $b$, respectively. Substituting these asymptotic expressions into the Friedmann equation
\begin{align}
 H_{\rm Full, eff}^2=\left(\frac{\dot{v}}{3v}\right)^2=\left(-\frac{{\rm sgn}(\gamma)}{3v\hbar}\frac{\partial {\cal H}^{\rm Full, eff}_{\rm tot}}{\partial b}\right)^2\,,
\end{align}
one can get the classical behavior of Hubble parameter as
\begin{align}\label{eq:H_eff2}
 H_{\rm Full, eff}^2\rightarrow\frac{8\pi G}{3}\left(\frac{g''(b_0)}{2}\,\rho_T+\frac{3\left[g'(b_0)\right]^2}{8\pi G\gamma^2\Delta}\right)\,.
\end{align}
For $b_0=b_{\rm c,I,1}=0$, we have
\begin{align}
  g''(b_0)=2,\quad g'(b_0)=0\,.
\end{align}
Hence the type-I-1 universe is asymptotically just the FRW universe. The asymptotic behavior of the type-I-2 universe depends on the values of $g'(b)$ and $g''(b)$ at $b=b_{\rm c,I,2}$. From Eq. \eqref{eq:H_eff2}, one can see that, if $g'(b_0=b_{\rm c,I,2})\neq0$, the type-I-2 universe would be an asymptotic de Sitter universe with a positive effective cosmological constant
\begin{align}\label{eq:eff-cosmological-constant}
\Lambda_{\rm eff}=\frac{3\left[g'(b_0=b_{\rm c,I,2})\right]^2}{\gamma^2\Delta}\,,
\end{align}
coupled to a scalar field with energy density
\begin{align}
\tilde{\rho}_T=\frac{g''(b_0=b_{\rm c,I,2})}{2}\,\rho_T\,.
\end{align} 
It turns out that for the case of $\gamma=0.2375$, the type-I-2 universe is an asymptotic de Sitter universe with
\begin{align}
 \Lambda_{\rm eff}=\frac{1.46635}{\Delta}\,,\quad \tilde{\rho}_T=0.90674\rho_T\,.
\end{align}
Hence, the quantum geometry in LQC contributes an effective cosmological constant.

\begin{figure}
  \includegraphics[width=\columnwidth]{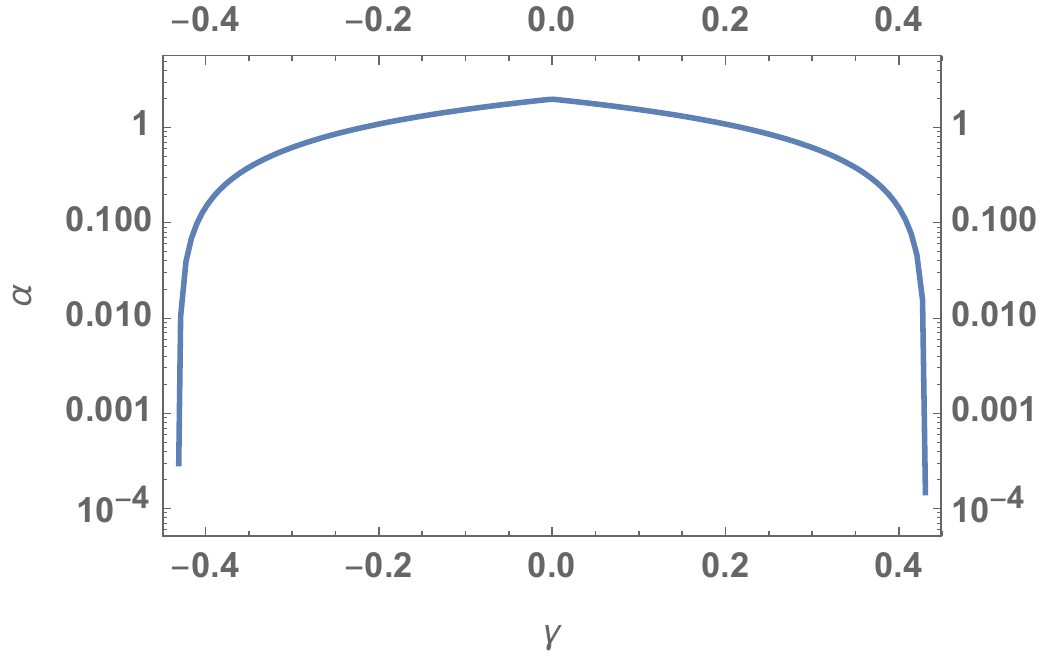}
    \includegraphics[width=\columnwidth]{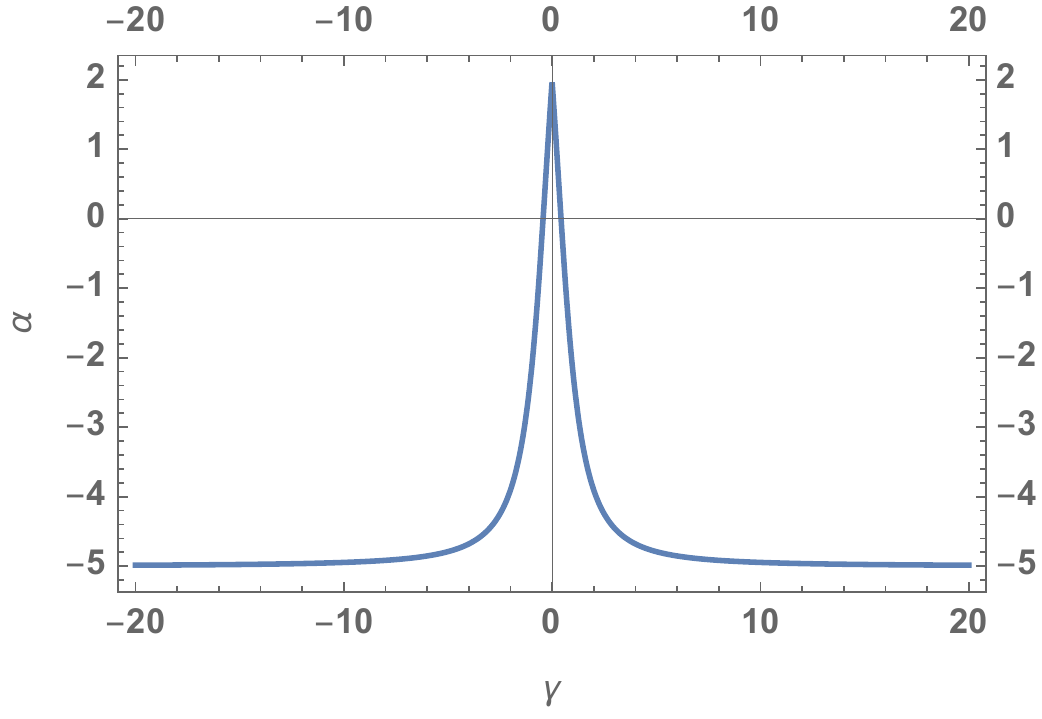}
  \caption{Plot of the factor $\alpha(\gamma)$ in Eq. \eqref{eq:Hubble-I-2}.} 
   \label{fig:alpha-gamma}
\end{figure}

\begin{figure}
  \includegraphics[width=\columnwidth]{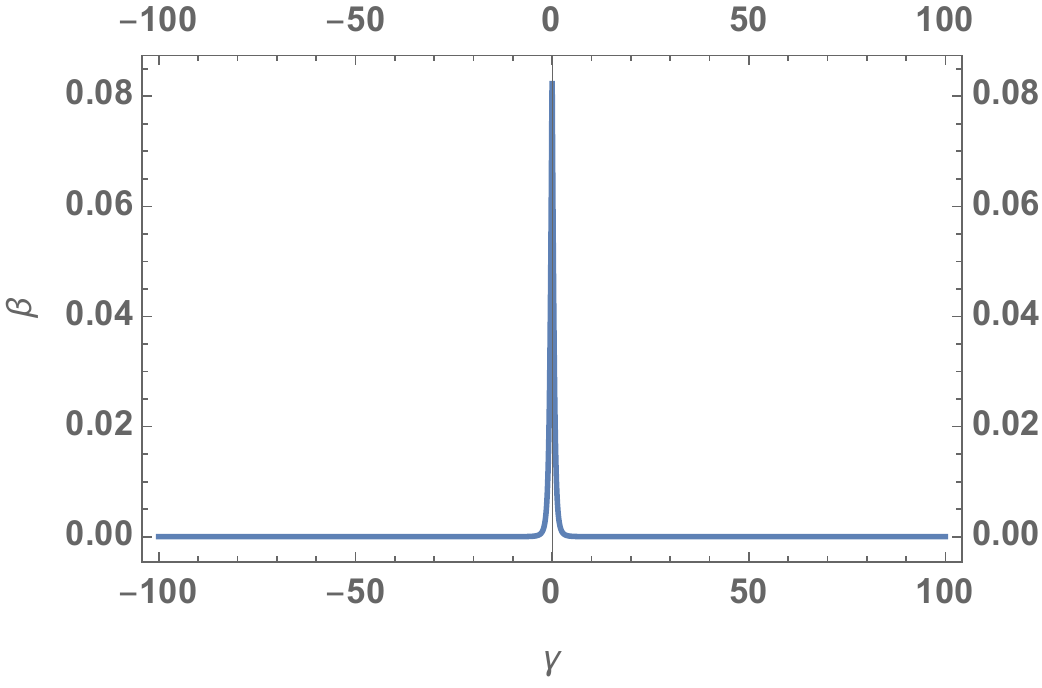}
  \caption{Plot of the factor $\beta(\gamma)$ in Eq. \eqref{eq:Hubble-I-2}.} 
   \label{fig:beta-gamma}
\end{figure}

To study how the energy density of scalar field and the cosmological constant in  the type-I-2 universe depend on $\gamma$, we write its effective Friedmann equation as the form
\begin{align}\label{eq:Hubble-I-2}
H^2_{\rm c,I,2}&=\frac{8\pi G}{3}\left[\alpha(\gamma)\,\rho_T+\frac{\beta(\gamma)}{G\Delta}\right]\,.
\end{align}
The factors $\alpha$ and $\beta$ in Eq. \eqref{eq:Hubble-I-2} as functions of $\gamma$ are plotted in \figref{fig:alpha-gamma} and \figref{fig:beta-gamma}, respectively. It turns out that $\alpha(\gamma)$ is positive for $\gamma\in(-0.430846,0.430846)$ and negative for $\gamma\in(-\infty,0.430846)\cup(0.430846,\infty)$, while $\beta(\gamma)$ is positive for all values of $\gamma$.  Therefore, the asymptotical FRW universe (the type-I-1 universe) will be bounced to an asymptotic de Sitter universe (the type-I-2 universe) coupled to a scalar field with positive or negative energy density for $\gamma\in(-0.430846,0.430846)$ or $\gamma\in(-\infty,0.430846)\cup(0.430846,\infty)$.

\begin{figure}
  \includegraphics[width=\columnwidth]{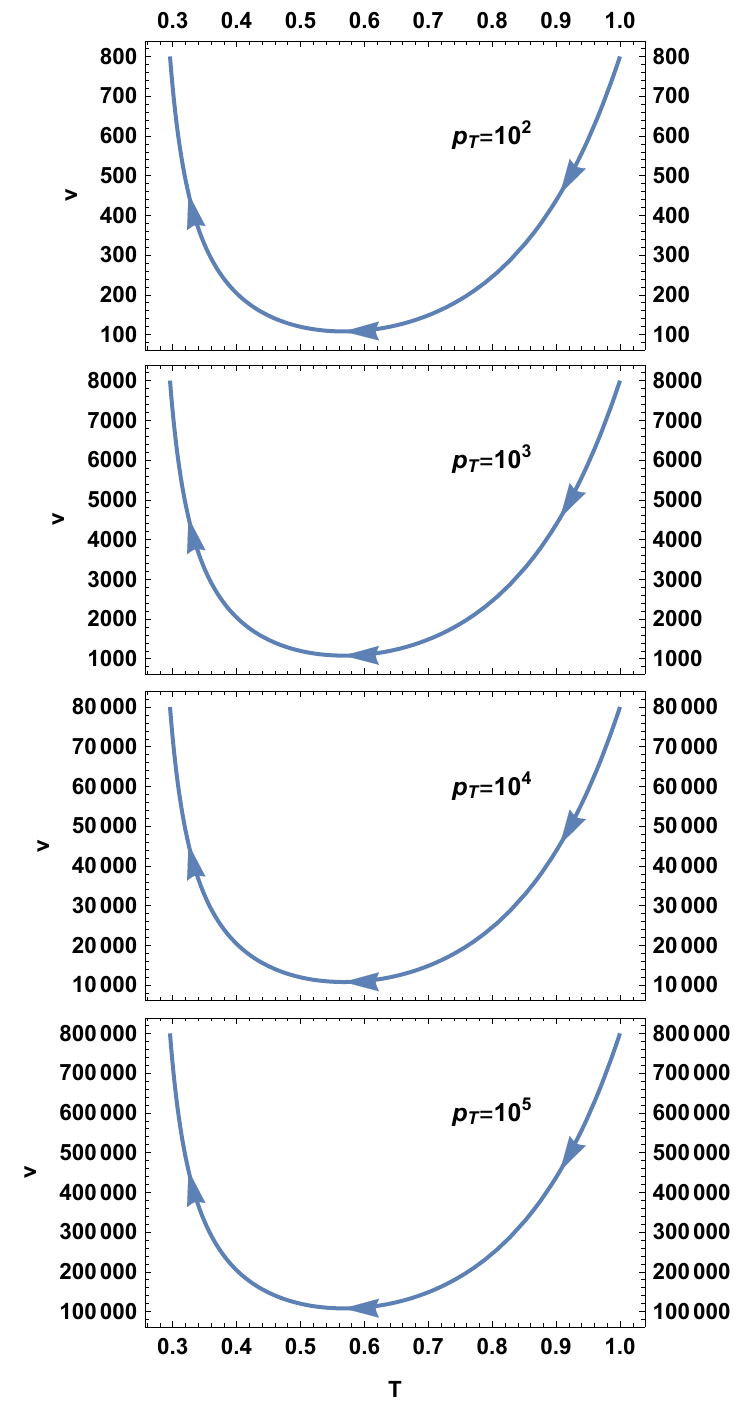}
  \caption{Plots of the relative evolution of $v$ with respect to the scalar field $T$, in reverse direction of the cosmological time,  given by the effective Hamiltonian constraint for different values of $p_T$ with initial data $v(0.0204175,p_T)=8p_T$ and $T(0.0204175,p_T)=1$, where $G=\hbar=1$, $\gamma=0.2375$.}
  \label{fig:phi-v}
\end{figure}

To confirm the above results, we can numerically investigate the relative evolution of $v$ with respect to $T$ determined by the effective Hamiltonian constraint \eqref{eq:Full-eff-Hamiltonian}. The constraint equation \eqref{eq:constraint-eq} yields
\begin{align}\label{eq:v-b}
v=v(b,p_T)=\frac{p_T}{2\sqrt{3\pi G\hbar^2 g(b)}}\,.
\end{align}
By Eq. \eqref{eq:v-b}, one can solve the evolution of $T$ with respect to $b$ as
\begin{align}\label{eq:evolution-T-b}
\frac{{\rm d}T}{{\rm d}b}=\frac{\{T,{\cal H}^{\rm Full,  eff}_{\rm tot}\}}{\{b,{\cal H}^{\rm Full,  eff}_{\rm tot}\}}=
-\frac{1}{{\rm sgn(\gamma)}2\sqrt{3\pi G g(b)}}\,.
\end{align}
Let the solution of Eq. \eqref{eq:evolution-T-b} be
\begin{align}\label{eq:b-pT}
T=T(b,p_T)\,.
\end{align}
Combining Eqs \eqref{eq:v-b} and \eqref{eq:b-pT} and then eliminating $b$ yield $v=v(T,p_T)$, which is plotted in \figref{fig:phi-v}. It shows that the classical big-bang singularity is again replaced by a quantum bounce. Moreover, the backward evolutions of all the dynamical solutions shown in \figref{fig:phi-v} are bounced from the FRW universe to asympotic de Sitter universes.

\section{Summary and discussion}
\label{sec-VI}

In the previous sections, the Euclidean Hamiltonian constraint operator proposed in \cite{Yang:2019ujs} was employed to quantize the Lorentzian term of the gravitational Hamiltonian constraint in the LQC model as in full LQG. A new alternative Hamiltonian constraint operator was obtained by the new regularization scheme for LQC. By the semiclassical analysis, we obtained the effective Hamiltonian constraint \eqref{eq:Full-eff-Hamiltonian} for the spatially flat FRW cosmology with a scalar field, which justified that the new quantum dynamics has the correct classical limit. It was shown that the quantum geometric correction to the Hamiltonian constraint is entirely encoded in the function $g(b)$ of Eq. \eqref{eq:g-b}. Note that the quantization scheme in Ref. \cite{Yang:2019ujs} contributed a function $g(b)$ different from \eqref{eq:g-b}. Both forms of $g(b)$ have the same classical limit $b^2$. Moreover, the energy density $\rho_T$ of matter can be related to $g(b)$ as Eq. \eqref{rho-relation} by the effective constraint equation, from which the critical matter density $\rho_{\rm crit}$ at the bounce point can be obtained. The existence of the critical density implied that the classical big-bang singularity could be resolved in the LQC model. 

The effective dynamics of the model was studied in details in Sec. \ref{sec-V}. It turns out that the physically possible values of the variable $b$ lie in three unconnected regions as shown in Fig. \ref{fig:g-v}. By the asymptotic analysis for $v\rightarrow\infty$, we found that the classical behavior of Hubble parameter is again determined by $g(b)$ as shown in Eq. \eqref{eq:H_eff2}. In the classical region I, there are two asymptotic solutions, $b=b_{\rm c,I,1}=0$ and $b=b_{\rm c,I,2}$, to $g(b)=0$, such that a prebounce branch (the type-I-1 universe) is bounced to a postbounce branch (the type-I-2 universe). The latter is an asymptotic de Sitter universe coupled to a massless scalar field with a positive effective cosmological constant $\Lambda_{\rm eff}=3\left[g'(b_0=b_{\rm c,I,2})\right]^2/(\gamma^2\Delta)$ and an energy density $\tilde{\rho}_T=\frac{g''(b_0=b_{\rm c,I,2})}{2}\,\rho_T$. It should be noticed that the energy density $\tilde{\rho}_T$ depends on $\gamma$, and it is positive for $\gamma\in(-0.430846,0.430846)$. Since the black hole entropy calculation in LQG indicates $\gamma=0.2375$ \cite{Domagala:2004jt,Meissner:2004ju}, our new quantization scheme of treating the Euclidean and Lorentzian terms independently for the Hamiltonian constraint can overcome the negative energy density problem in \cite{Yang:2019ujs}, while the asymmetric bounce still holds.

It is interesting to notice that if one adopted a massless scalar field with negative energy in the universe as in \cite{Bekenstein:1975ww,Narlikar:1986dc}, it would be possible to predict a positive cosmological constant of our universe matching up the observation. To see this, recall that the observation indicates the cosmological constant $\Lambda_{\rm obs}\approx 1.09168\times 10^{-52}\; {\rm m^{-2}}$ \cite{Aghanim:2018eyx}. To match the effective cosmological constant up $\Lambda_{\rm obs}$, Eq. \eqref{eq:eff-cosmological-constant} implies a large value of $\gamma$, which would give a negative energy density of the scalar field.
\begin{acknowledgments}
J. Y. would like to thank Professor Chopin Soo for useful discussions. This work is supported in part by NSFC Grants No. 11765006, No. 11875006, No. 11961131013. C. Z. acknowledges the support by the Polish Narodowe Centrum Nauki, Grant No. 2018/30/Q/ST2/00811.
\end{acknowledgments}


%

\end{document}